\documentclass
[twocolumn,showpacs,preprintnumbers,amsmath,amssymb]
{revtex4}
\usepackage{graphicx}% Include figure files
\usepackage{dcolumn}% Align table columns on decimal point
\usepackage{bm}% bold math
\usepackage{here}
\usepackage{color}
\usepackage{braket}
\usepackage{ascmac}

\begin{document}

%% \u30d9\u30af\u30c8\u30eb\u3001\u5927\u5206\u6570\u30b3\u30de\u30f3\u30c9
\newcommand{\vc}[1]{\mbox{\boldmath $#1$}}
%% \u5206\u6570
\newcommand{\fracd}[2]{\frac{\displaystyle #1}{\displaystyle #2}}
%% \u8d64\u8272
\newcommand{\red}[1]{\textcolor{red}{#1}}
\newcommand{\blue}[1]{\textcolor{blue}{#1}}
\newcommand{\green}[1]{\textcolor{green}{#1}}
%% \u504f\u5fae\u5206

%\def\red{\textcolor{red}}
%\def\blue{\textcolor{}}
%\def\green{\textcolor{green}}

%\newcommand{\del}{\partial}

\def\ni{\noindent}
\def\nn{\nonumber}
\def\bH{\begin{Huge}}
\def\eH{\end{Huge}}
\def\bL{\begin{Large}}
\def\eL{\end{Large}}
\def\bl{\begin{large}}
\def\el{\end{large}}
\def\beq{\begin{eqnarray}}
\def\eeq{\end{eqnarray}}
\def\beqnn{\begin{eqnarray*}}
\def\eeqnn{\end{eqnarray*}}

\def\bit{\begin{itemize}}
\def\eit{\end{itemize}}
\def\bsc{\begin{screen}}
\def\esc{\end{screen}}

%%short-handed
\def\eps{\epsilon}
\def\th{\theta}
\def\del{\delta}
\def\omg{\omega}

\def\e{{\rm e}}
\def\exp{{\rm exp}}
\def\arg{{\rm arg}}
\def\Im{{\rm Im}}
\def\Re{{\rm Re}}

\def\sup{\supset}
\def\sub{\subset}
\def\a{\cap}
\def\u{\cup}
\def\bks{\backslash}

\def\ovl{\overline}
\def\unl{\underline}

\def\rar{\rightarrow}
\def\Rar{\Rightarrow}
\def\lar{\leftarrow}
\def\Lar{\Leftarrow}
\def\bar{\leftrightarrow}
\def\Bar{\Leftrightarrow}

\def\pr{\partial}

%\カッコ
\def\>{\rangle} %>
\def\<{\langle} %<
\def\RR {\rangle\!\rangle} %>>
\def\LL {\langle\!\langle} %<< 
\def\const{{\rm const.}}

\def\e{{\rm e}}

\def\Bstar{\bL $\star$ \eL}

%%%%%%%%%%%%%%%%%%%%%%%%%%%
%%%LOCALDEFLOCALDEFLOCALDEF
%%%%%%%%%%%%%%%%%%%%%%%%%%%
% LQPU
\def\etath{\eta_{th}}
\def\irrev{{\mathcal R}}
%%%%%%%%%%%%%%%%%%%%%%%%%%%
\def\e{{\rm e}}
\def\noise{n}
%\textcolor{blue}{example}
%%%LOCALDEFLOCALDEF%%%%
%%% hat %%%%
\def\hatp{\hat{p}}
\def\hatq{\hat{q}}
\def\hatU{\hat{U}}

%%% hatつき文字 %%%%
\def\hatA{\hat{A}}
\def\hatB{\hat{B}}
\def\hatC{\hat{C}}
\def\hatJ{\hat{J}}
\def\hatI{\hat{I}}
\def\hatP{\hat{P}}
\def\hatQ{\hat{Q}}
\def\hatU{\hat{U}}
\def\hatW{\hat{W}}
\def\hatX{\hat{X}}
\def\hatY{\hat{Y}}
\def\hatV{\hat{V}}
\def\hatt{\hat{t}}
\def\hatw{\hat{w}}

\def\hatp{\hat{p}}
\def\hatq{\hat{q}}
\def\hatU{\hat{U}}
\def\hatn{\hat{n}}
\def\hatb{\hat{b}}
\def\hata{\hat{a}}

\def\hatphi{\hat{\phi}}
\def\hattheta{\hat{\theta}}

%%%%LOCALDEFLOCALDEFLOCALDEF%%%%
\def\iset{\mathcal{I}}
\def\fset{\mathcal{F}}
\def\pr{\partial}
\def\traj{\ell}
\def\eps{\epsilon}
\def\U{\hat{U}}
%%%%%%%%%%%%%%%%%%%%%%%%%%%%%%%%

%%LOCALDEFLOCALDEF%%%
\def\U{U_{\rm cls}}
\def\P{P_{{\rm cls},\eta}}
\def\traj{\ell}
\def\cc{\cdot}

%LOCALDEFLOCALDEFLOCALDEF
\def\DZ{D^{(0)}}
\def\Dcls{D_{\rm cls}}

%LOCALDEFLOCALDEFLOCALDEFLOCALDEF
\def\alphains{\alpha_{ins}}
%\def\alphains{\alpha}
%LOCALDEFLOCALDEFLOCALDEFLOCALDEF

\newcommand{\relmiddle}[1]{\mathrel{}\middle#1\mathrel{}}

%Dynamical Anderson Transition in Kicked Models
\title{Localized-Diffusive and Ballistic-Diffusive Transitions in Kicked Incommensurate lattice
%Localization and delocalization properties in quasi-periodically 
%perturbed kicked Harper  model
}
%\title{Localization-Delocalization Transition in 
%Dynamically Perturbed 1D Disordered System}
%\title{Critical Phenomena of Dynamical Delocalization in Quantum Anderson Map}
%\title{A Numerical Test of Pade Approximation for Some Functions}
\author{Hiroaki S. Yamada}
%\email{hyamada[at]uranus.dti.ne.jp}
\affiliation{Yamada Physics Research Laboratory,
Aoyama 5-7-14-205, Niigata 950-2002, Japan}
%%\author{Fumihiro Matsui} 
%\email{rp00863[at]ed.ritsumei.ac.jp}
%%\affiliation{Department of Physics, College of Science and Engineering, Ritsumeikan University
%%Noji-higashi 1-1-1, Kusatsu 525-8577, Japan}
\author{Kensuke S. Ikeda}
%\email{ahoo[at]ike-dyn.ritsumei.ac.jp}
%}
\affiliation{College of Science and Engineering, Ritsumeikan University, 
Noji-higashi 1-1-1, Kusatsu 525-8577, Japan}

\date{\today}
\begin{abstract}
%準周期系は一般に局在相と弾道相を持つ。典型的準周期系であるKHMを用いて
%準周期系の局在相および弾道相への動的なコヒーレント摂動の効果が探求された。
%局在相からは$t^\alpha$($0<\alpha<1$)なる異常拡散を経て非可逆的な
%拡散層に転移しこれは従来局在乱雑格子でみられたものであるが、
%弾道相でも１より大きい指数の異常拡散を経て拡散層がおき、
%弾道-拡散転移の存在が初めて確かめられた。
By using the kicked Harper model, the effect of dynamical perturbations
to the localized and ballistic phases in aperiodic lattice systems is 
investigated. The transition from the localized phase to diffusive phase
via a critical sub-diffusion $t^\alpha$($t$:time) with $0<\alpha<1$ 
is observed. In addition, we first confirmed the existence of the 
transition from the ballistic phase to the diffusive phase via 
a critical super-diffusion with $1<\alpha<2$.
\end{abstract}

%05.45.Mt Quantum chaos; semiclassical methods
%71.23.Anderson localization: disordered solids
%72.20.Ee Mobility edges
%%%%%%%%%%%%%%%%%%
%%%%%%%%%%%%%%%%%%
%%%%%%%%%%%%%%%%%%
%03.00.00 Quantum mechanics, field theories, and special relativity
%03.65.-w Quantum mechanics
%03.75.-b Matter waves
%%%%%%%%%%%%%%%%%%
%05.00.00 Statistical physics, thermodynamics, and nonlinear dynamical systems
%05.60.Gg Quantum transport
%05.30.-d Quantum statistical mechanics
%%%%%%%%%%%%%%%%%%
%71.00.00 Electronic structure of bulk materials
%71.23.An Theories and models; localized states
%%%%%%%%%%%%%%%%%%
%72.00.00 Electronic transport in condensed matter
%72.10.-d Theory of electronic transport; scattering mechanisms
%72.10.Bg General formulation of transport theory
%72.10.Di Scattering by phonons, magnons, and other nonlocalized excitations
%72.20.Ee Mobility edges; hopping transport
%72.80.Ng Disordered solids
%%%%%%%%%%%%%%%%%%
%73.00.00 Electronic structure and electrical properties of surfaces, 
%interfaces, thin films, and low-dimensional structures
%73.43.Cd Theory and modeling

%\pacs{05.45.Mt,03.65.-w,05.30.-d}
% insert suggested keywords - APS authors don't need to do this
%\keywords{}

%\maketitle must follow title, authors, abstract, \pacs, and \keywords
\maketitle

% body of paper here - Use proper section commands
% References should be done using the \cite, \ref, and \label commands
%\section{}
% Put \label in argument of \section for cross-referencing
%\section{\label{}}
%\subsection{}
%\subsubsection{}

%%%%%%%%%%%%%%%%%%%
%%%%%%%%%%%%%%%%%%%
%\section{Introduction}
%\label{sec:intro}
%%%%%%%%%%%%%%%%%%%
%%%%%%%%%%%%%%%%%%%
{\it Introduction.-}
%%%%%%%%%%%%%%%%%%%%%%%%%%%
In 1D tight-binding model with any disorder the dynamical quantum states 
are always localized \cite{ishii73}
%Of course,when coupled
But if it is coupled with noisy source such as irregular thermal lattice vibrations,
the localization is immediately destroyed and is taken the place 
by normal diffusion \cite{haken72,haken73}.
%begins to spread and normal diffusion occurs.
%However, 
Even very simple time-dependent harmonic perturbations consisting 
only a few number of periods may convert the localized wave packet
into irreversible diffusion in Anderson model and kicked Anderson model (Anderson map)
\cite{yamada18,yamada20,yamada22}.
%
%cause the irreversible diffusion 
%in the localized wave packet.
The characteristic feature of such coherent perturbations 
is that it realizes irreversible diffusion  %occurs 
from the localized state via a critical phase transition 
we call localization-delocalization transition (LDT).

1D aperiodic lattice systems have the both features of disordered lattice and periodoic lattice.
Indeed it has ballistic states and critical states in addition to the localized states.
%A remarkable feature of 1D aperiodic lattice systems is that in addition to the localized 
%state there exist ballistic states and critical states.
% ballistic states.
Our question arising here is what would be the effect of the coherent perturbations 
on the dynamical states of 1D aperiodic lattice systems, i.e.., 
the localized states, the critical states and the ballistic states 
%in the 1D aperiodic lattice systems, 
%and can the states induce the normal diffusion?

To answer the above question, we investigate the effects of the coherent perturbation
by using kicked Harper model (KHM) that facilitate an efficient study 
of the dynamics in aperiodic lattice. In fact, the KHM shows the above three states
intrinsic localized, critical, and ballistic states.
The first question is whether the LDT observed for the coherently perturbed random system
also exists in the aperiodic lattice system.
If this is the case, considering the remarkable duality property of the KHM,  
a transition from the ballistic to the diffusive state is also expected.
The purpose of the present paper is to confirm above conjectures.
%Is the transition also induced by the coherent perturbation?

 As stated above, there are three types of main dynamical states of the quantum wave packet, 
localized, normal diffusion, and ballistic spread,
%\red{motionは？spreadというのが以下慣用句だが業界用語？）}, 
in the KHM depending upon the potential strength. 
In that respect, it is the same as the Harper model without the kicks.
The phase diagram of the localized/delocalized state in KHM 
 has a nested structure and is quite complicated
 \cite{artuso94,prosen01,kolovsky03,levi04}.
%In the case of KHM, there is a study on the LDT that occurs 
% when the second kick series or kick intensity modulation is applied. 
As for the LDT, its presence was reported by some authors for the KHM
driven by the 
doublly periodic kick or the intensity modulated periodic kicks.
 %●
%Recently, the LDT by using the KR has been
%experimentally explored using cold atoms in optical lattices
%\cite{sadgrove07,dana08,chabe08,lemarie10}. 
It is also feasible to experimentally  observe the dynamical LDT
through the diffusion of wave packets
in the pulsed 1D incommensurate optical lattice such as the KHM
\cite{sarkar17}.

The purpose of this paper is to show realization of the normal diffusion by simple
harmonic perturbations via the above two transitions paths, where the harmonic perturbation
is polychromatic and is composed of $M$ frequencies mode.
%\blue{なぜこの文があるのか？
%The normal diffusion can easily occur even in tight-binding systems 
%without temporal noise or spatial randomness. }\\
%
%\blue{以下の文はKHMの局在相に対していってるのか？ 誤解招く心配?\\
%In the monochromatically perturbed case $M=1$  is localized. 
%The perturbed KHM shows LDT above $M\geq2$ as increase of
% the perturbation strength $\eps$.
% The localized state exhibits transition to the irreversible diffusive
%state through an anomalous diffusion  $t^\alpha$ with the critical exponents $0<\alpha_c <1$
%similar to the Anderson transition. }\\
%以上の意を勘案しての文案\\
We first show the localized state of KHM responds to such perturbation in quite similar 
ways to that of kicked disordered lattice systems such as 
the Anderson map \cite{yamada18,yamada20} and 
quantum standard map \cite{casati89,delande08,delande13,yamada20}: 
for monochromatic perturbation $M=1$ the localizaion still remains and it is $M\geq2$ 
that the KHM localized state undergoes the LDT with increase in the perturbation strength $\eps$.
We next show the response of the extended (ballistic) state of KHM to the harmonic perturbation.
We report the wave packet dynamics exhibiting the ballistic spreading changes into 
the normal diffusion when the perturbation strength is increased. The transition
occurs through a novel anomalous diffusion as the critical state, which
is characterized by a ``super'' diffusion exponent  $1<\alpha_c <2$. 
%It is shown that a dynamical transition from ballistic to diffusive spread. 
Hereinafter, this transition is referred to as a ballistic-diffusive transition (BDT) 
of the wave packet spreading.

%量子局在系に周期的コヒーレントな動的自由度を印加すると自由度があたかも空間次元の
%役割を果たし数自由度で非局在化転移が起きてtime-irreversible な正常拡散が発生する。
%potential にrandomness が含まれず、それゆえ　potential energy とkinetic ennergy
%の相対的大きさに依存してballistic、localizing を共に示すamphibious な準周期系でも
%coherent 自由度の印加によって同様な正常拡散への臨界転移が観測されるか否かが本論
%文の解明目的。

%When periodic coherent dynamical degrees of freedom are applied to
% a quantum localized system, the degrees of freedom play the role 
% of spatial dimensions, and a delocalization transition occurs 
% with a few degrees of freedom, resulting in time-irreversible normal diffusion.
%One of the objectives of this paper is to investigate whether the similar critical transition 
%to normal diffusion can be observed by applying coherent perturbation to the KHM 
%that do not include randomness in the potential and exhibit both ballistic and localizing states
% depending on the relative magnitudes of potential energy and kinetic energy.

%%%%%%%%%%%%%%%%%%%
%%%%%%%%%%%%%%%%%%%
%\section{Models and some preliminaries}
%\label{sec:model}
%%%%%%%%%%%%%%%%%%%
%%%%%%%%%%%%%%%%%%%
{\it Model.-}
%%%%%%%%%%%%%%%%%%%%%%%%%%%
We deal with  the dynamically perturbed kicked Harper model (KHM),
%$i\partial_t u(n,t) = u(n-1,t)+u(n+1,t)+f(t)V(n)u(n,t), $
\beq
H(t)&=&J\sum\limits_{n}^N (\hatb_n^\dagger \hatb_{n+1}+H.C.) \nn \\
 &+& 
2V\left[1+f_\eps(t)\right] \delta_1(t)\sum\limits_{n}^N\cos(2\pi Q n)\hatb_n^\dagger\hatb_n,
\label{eq:Hamiltonian}
\eeq
where $\hatb_n$($\hatb_n^\dagger$) is the creation(annihilation) operator of the 
n-th site, and $N$ is the lattice size.
$\delta_1(t)=\sum_{m\in {\Bbb Z}}\delta(t-m)$.
%The on-site energy sequence is 
%\beq
 % V(n)=2V \cos(2\pi Q n), 
%\eeq
%時間摂動を次のようにとる。($m_j\in {\Bbb Z}$)
%where $\{|n \rangle \}$ is an orthonormalized basis set and 
The $Q(=\frac{\sqrt{5}-1}{2})$ is an irrational number. 
 $V$ and $J(=-1)$ denote the potential strength and the hopping energy 
 between adjacent sites, respectively.
%We tale $Q=\frac{\sqrt{5}-1}{2}$ and $T=-1$ throughout the present paper.

The harmonic perturbation $f_\eps(t)$ is the sum of harmonic functions
 \beq
  f_\eps(t)=\eps f(t)=\frac{\eps}{\sqrt{M}} \sum_i^M\cos(\omega_i t), 
%    f(t)=\frac{\eps}{\sqrt{M}} \sum_i^M\cos(\omega_i t + \theta_i), 
\eeq
where $M$ and $\eps$ are the number of frequency components and 
the relative strength of the perturbation, respectively.
The frequencies $\{ \omega_i\}(i=1,...,M)$ are taken as mutually incommensurate 
numbers of order $O(1)$.
Note that the long-time average of the total power of the perturbation is normalized to 
$\overline{f_\eps(t)^2}=\eps^2/2$.

In the spatial continuous limit, $H(t)$ is given by 
\beq
H_c(t)=2J\cos(\hatp/\hbar)+2V\cos(2\pi Q \hatq) [1+f_\eps(t)]\delta_1(t), 
\eeq
where $\hatp$ and $\hatq$ denote the momentum 
and position operators, respectively.
The unperturbed KHM ($\eps=0$) is known to 
take a localized state ($|V|>>|J|$), critical state ($V=J$), and extended ($|V|<<|J|$) state
%\red{これはballistic stateの方がよかない？} 
depending on $V$.

The initial wave packet $<n|\Psi(t=0)>=\delta_{n,N/2}$ is localized at the site $N/2$, and
we calculate the time evolution of the wavefunction $|\Psi(t)>$ 
using  Schr\''{o}dinger equation. 
We monitor the spread of the wave function in the site space by the 
mean square displacement (MSD):
$m_2(t) = \sum_{n}(n-N/2)^2 \left< |\phi(n,t)|^2 \right>$, 
where $\phi(n,t)=<n|\Psi(t)>$ is the site representation of the wave function.
The number of steps is $10^5\sim10^6$.
We mainly use the system size $N=2^{16}-2^{17}$, 
and $\hbar=1/8$.

At the critical state of the transition, an anomalous diffusion $m_2(t) \sim t^\alpha$
% (0<\alpha<2)$ これは余計なので消す
characterized by diffusion index $\alpha$ is expected.
% at the trnsition point.
To observe such a behavior directly we introduce time-dependent
the instantaneous diffusion index $\alphains(t)$ defined by
$\alphains(t):=\frac{d\log\ovl{m_2(t)}}{d\log t}$, where
the locally time-averaged MSD $\ovl{m_2(t)}$ is used in order to reduce 
fluctuation in a shorter time-scale.

%is also used 
%to directly investigate the existence of the LDT:
%Above the critical point $\eps>\eps_c$ the index becomes unity 
%($\alpha=1$) indicating the normal diffusion,
%and for $\eps<\eps_c$ it decreases to zero ($\alpha=0$)  indicating localization.
%Around the critical point $\eps \simeq \eps_c$  for LDT
% $\alpha_{ins}(t) \simeq \alpha_c$.

\begin{figure}[htbp]
\begin{center}
\hspace{-8mm}
\includegraphics[width=4.5cm]{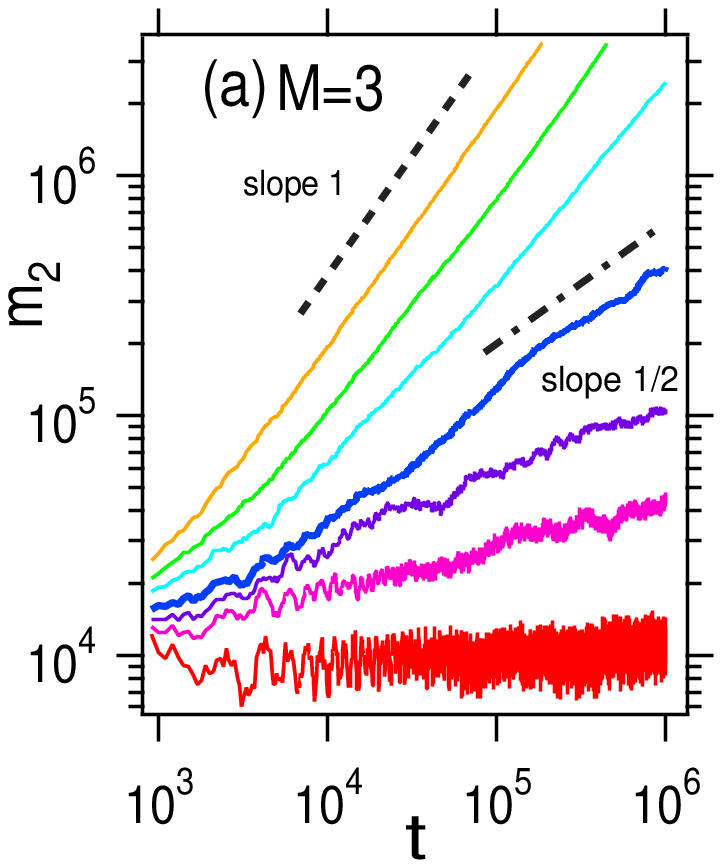}
\hspace{-5mm}
\includegraphics[width=4.5cm]{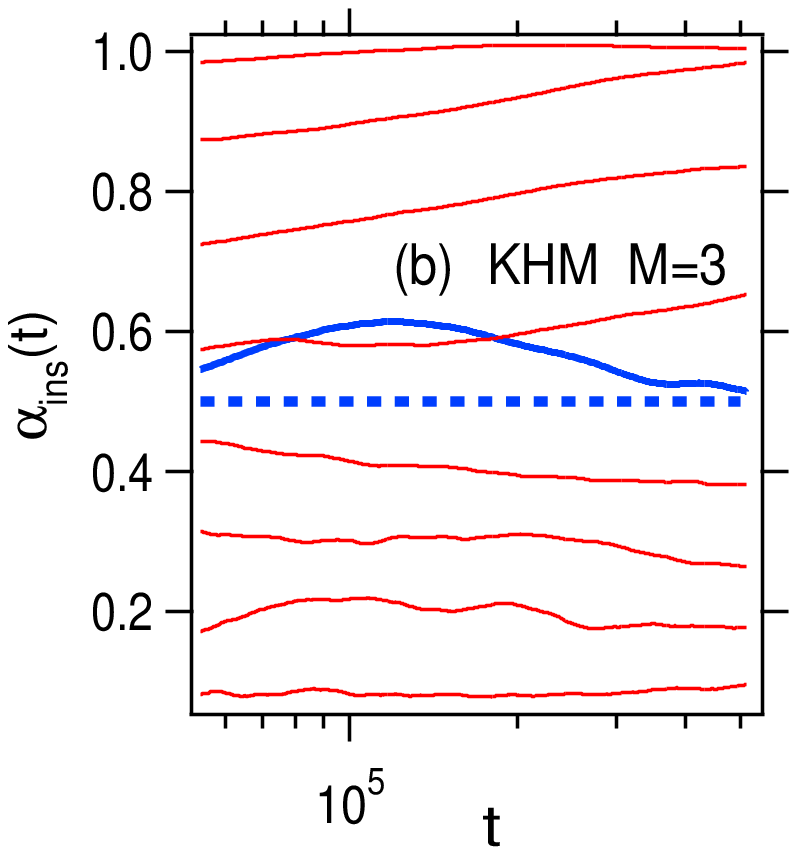}
\hspace{8mm}
\includegraphics[width=5.5cm]{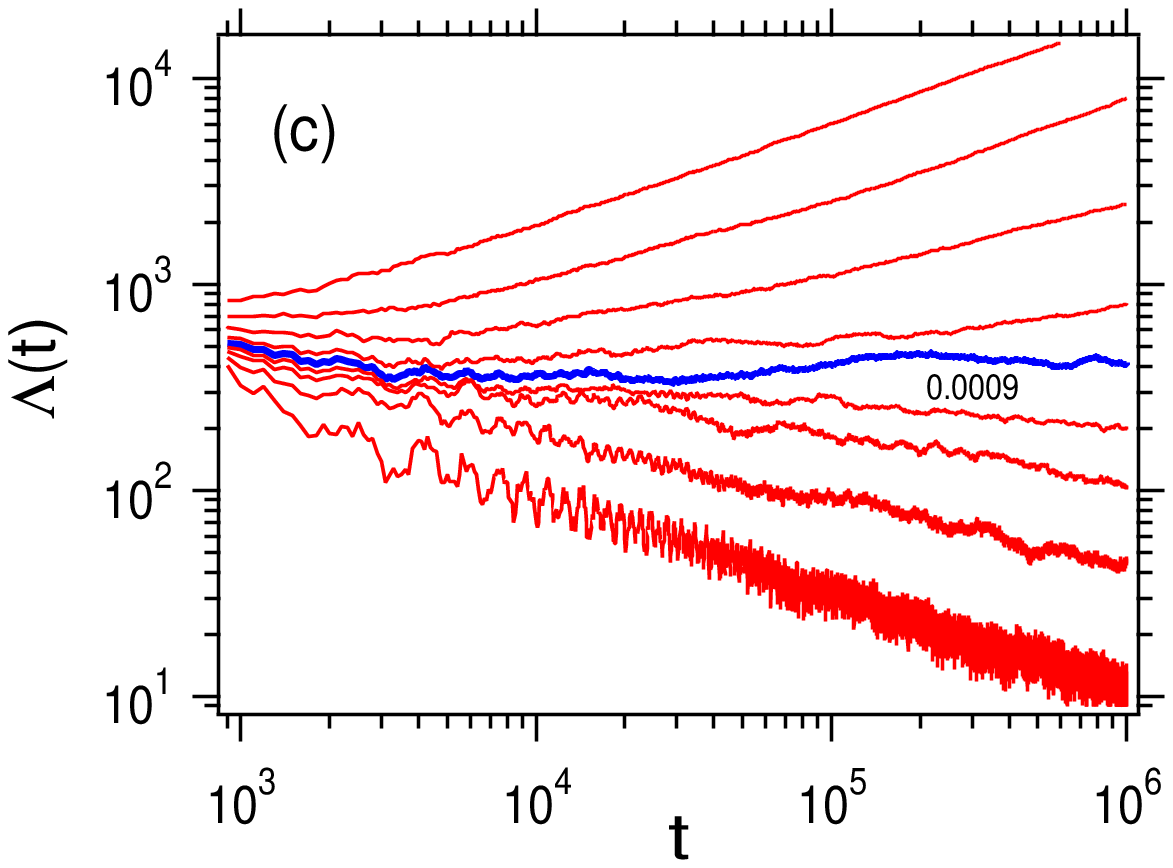}
\caption{(Color online) \label{fig:LDT}
 (a)The double-logarithmic plots of $m_2(t)$ as a function of time
for different values of the perturbation strength $\eps$
 increasing from $\eps=0.0001$(bottom)  to $\eps=0.002$(top)
in  the polychromatically perturbed  
KHM of $V=5$ with $M=3$.. $\hbar=1/8$.
%The values of $\eps$ used are $\eps=0.0002,0.0020,0.0040,0.0080,0.0200,0.1800$ 
%from bottom to top.
(b)The instantaneous diffusion index $\alpha_{ins}(t)$ for some $\eps$.
The broken line  indicates the critical sub-diffusion line
$\alpha_{ins}(t)=\alpha_c=0.5$ predicted by the scaling theory.
 (c) The scaled MSD $\Lambda(\eps,t)=m_2(t)/t^{1/2}$ 
as functions of time for increasing perturbation strengths. 
The results for $\eps_c \simeq 0.0009$ are shown in thick blue lines.
%The range is $\eps \in[0.0001, 0.002]$.
%$\eps=0.17,0.18,0.19,0.20,0.21,0.22,0.23,0.25$ from below.
}
\end{center}
\end{figure}

%%%%%%%%%%%%%%%%%%%%%%%%%%%%
{\it Localization-delocalization transition (LDT).-}
%%%%%%%%%%%%%%%%%%%%%%%%%%%
%\blue{
We investigate the delocalization in the KHM with $V=5$.
%\red{○池田案
%イキナリ $M=3$はキツい。たとえばこんなことは言えないか？ 
%ballisticの場合はM=1はダメの言い訳がある
In the case of $M=1$ the localization maintains for the harmonic perturbaion,
and for $M\ge2$ the  delocalization is observed on a finite time scale, 
which agrees with the case of kicked Anderson model \cite{yamada20}.
Here we focus on the case $M=3$. 
%}
In Fig.\ref{fig:LDT}(a), the MSD are shown for $M=3$.
They indicate localization when $\eps$ is small, but the LDT occurs at
 a certain critical value $\eps_c$, and $\eps$ exceeds $\eps_c$, the plot of MSD  
 warping upward (upward deviation) can be seen  in the double-logarithmic plots.
For $\eps>\eps_c$ the normal diffusive behavior $m_2 \sim t^1$ appears as $t \to \infty$.
Around $\eps=\eps_c$, 
the time-dependence of  MSD can be approximately described 
by the sub-diffusive spreading.
As a result, the time-dependent MSD displays opposite behaviors for $\eps<\eps_c$
and for $\eps>\eps_c$, which can directly be confirmed
by $\alphains(t)$ vs $t$ plot.
As shown in Fig.\ref{fig:LDT}(b), 
in the case of $V>>1$, we can see that there exist an $\eps_c$ above which the
plot of $\alphains(t)$ increases and below which it decreases.
The limiting tendency of $\alphains(t)$ for $t\to\infty$  seems to be
$\alphains \to 1$ or $\alphains(t)\to 0$ for $\eps>\eps_c$ or for $\eps>\eps_c$, 
respectively. At $\eps=\eps_c$,~$\alphains(t)$ fluctuates around $\alphains(t)\sim 1/2$,
which indicate the anomalous diffusion of $m_2\propto t^{\alpha_c}$ with $\alpha_c\sim 0.5$.

As shown in Fig.\ref{fig:LDT}(c), the scaled MSD $\Lambda(t)=m_2(t)/t^{1/2}$ 
for various $\eps$ has trumpet-shaped pattern 
that suggests the existence of the LDT \cite{yamada20}.

%%%%%%%%%%%%%%%%%%%%%%%%%%%%%%%%%%%%%%%%%%
%(a)M3b.pxp/V5M3/V5-loc-eps-tuika-letter/sin-KHM-A-Loc-eps
%(c)M3b-scale.pxp/V5M3/V5-loc-eps-tuika-letter/sin-KHM-A-Loc-eps
%(b)c3b-alpha.pxp/c3/alpha-Atype-KHM-LDT/sin-KHM-A-Loc-eps
%%%%%%%%%%%%%%%%%%%%%%%%%%%%%%%%%%%%%%%%%%

\begin{figure}[htbp]
\begin{center}
\includegraphics[width=6.0cm]{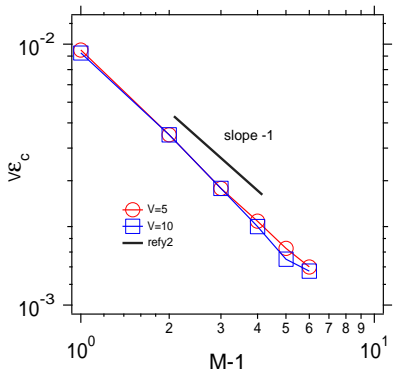}
\caption{\label{fig:V5epsc-M}(Color online)
The critical perturbation strength $V \eps_c$ as a function of  $(M-1)$ 
for $V=5$ and $V=10$ in the KHM. 
The black solid line with  slope $-1$ is shown as a reference.
%shows $\eps_c \propto 1/(M-1)$.
}
\end{center}
\end{figure}

As shown in Fig.\ref{fig:V5epsc-M}, for $M\geq 2$ the $M-$dependence of the 
critical strength $\eps_c$ indicates the inverse power-law 
\beq
 \eps_c \propto \frac{1}{V}\frac{1}{(M-1)}.
\eeq
The same $M-$dependence even for  $V=10$ can be obtained,
as seen in Fig.\ref{fig:V5epsc-M}.
This difference in $\eps_c$ due to $V$ can be interpreted by the Maryland transform
 in appendix. 
According to the Maryland transform, the effect of $V$ in the diagonal term is saturated 
in the region $V>V^*(\equiv 0.38)$ for $\hbar=1/8$, and 
 the nature of the off-diagonal term depends on $\eps V$, so the critical value $\eps_c$ 
  for $V=10$ can be interpreted as a half of the transition point for $V=5$. 

%\blue{
We plotted in Fig.\ref{fig:V5epsc-M} dependence of the 
substantial critical perturbation strength $\eps_c V$ 
upon the color number, which leads to
\beq
 \eps_cV  \propto \frac{1}{(M-1)}.
\eeq
An interesting fact is that the result means that
$\eps_c V$(more precisely $\eps_c V/\hbar$) does no longer
dependent upon $V$ for $|V| \gg 1$. It implies the effect 
of the quasi-periodic perturbation on the localization
effect saturates as $|V|$ increases. This can immediately 
be understood by the Maryland transformed scheme
of the quasi-energy eigenvalue equation 
for the time-periodic Hamiltonian (\ref{eq:Hamiltonian}), 
which is shown in Appendix. 
It is interesting that the LDT occurs at a very small relative amplitude
of perturbation, ie, $\eps_c\sim 10^{-3}$, in contrast to the BDT to be given later. 
%}

%%%%%%%%%%%%%%%%%%%%%%%%%%%%%%%%%%%%%%%%%%
%HarperMap確認日歯M123
%V5epsc-M.pxp/critical-epsc  hbar=1/8 theta=0
%%%%%%%%%%%%%%%%%%%%%%%%%%%%%%%%%%%%%%%%%%

\begin{figure}[htbp]
\begin{center}
\hspace{-8mm}
\includegraphics[width=4.5cm]{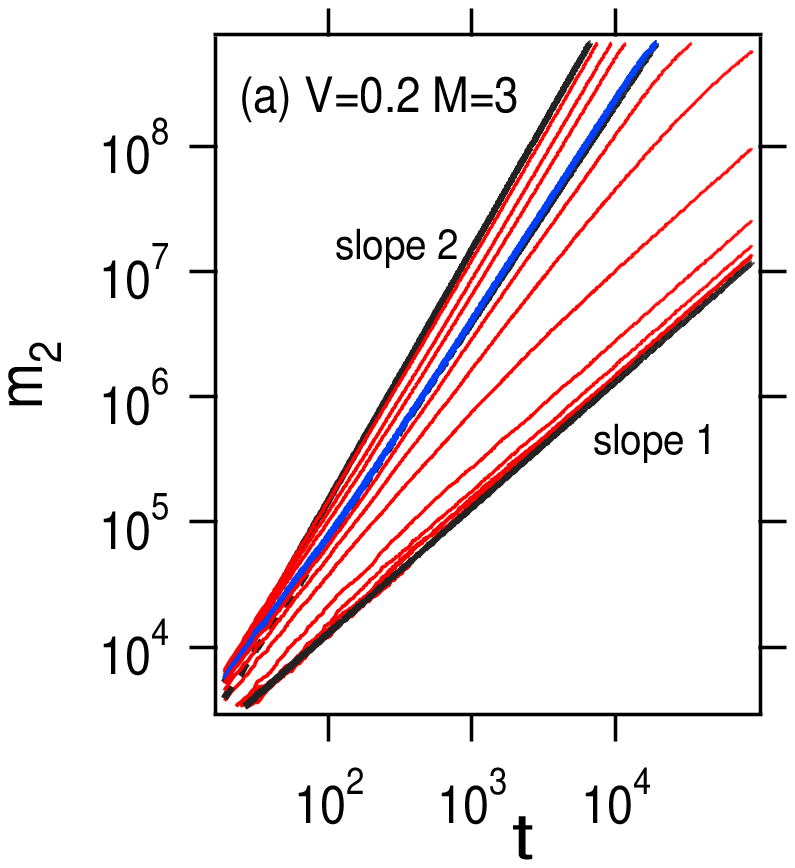}
\hspace{-5mm}
\includegraphics[width=4.5cm]{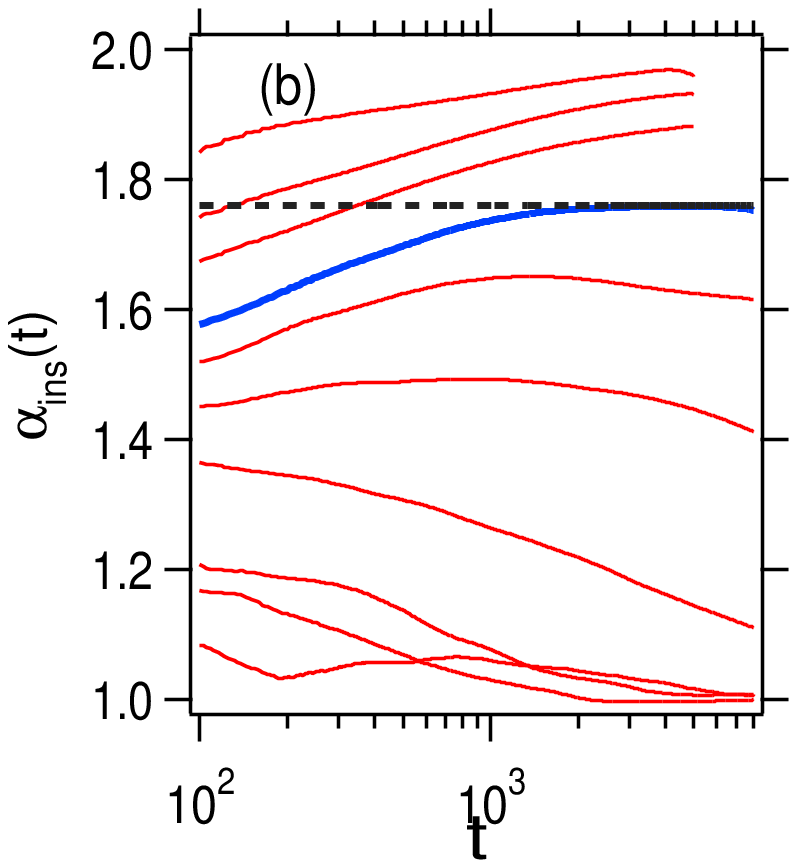}
\hspace{8mm}
\includegraphics[width=5.5cm]{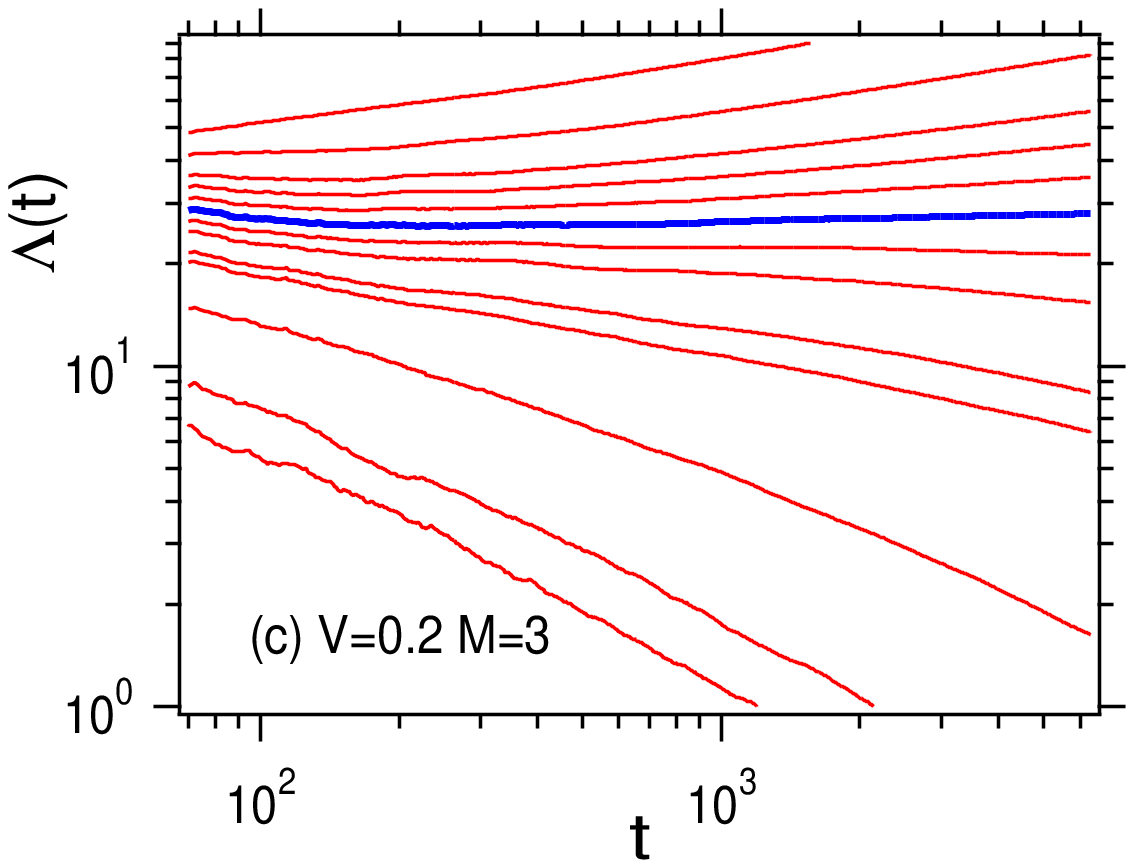}
\caption{(Color online) \label{fig:BDT}
 (a)The double-logarithmic plots of $m_2(t)$ as a function of time
for different values of the perturbation strength $\eps$
 increasing from $\eps=0.02$(top)  to $\eps=0.8$(bottom)
in  the polychromatically perturbed  
KHM of $V=0.2$ with $M=3$. $\hbar=1/8$.
%The values of $\eps$ used are $\eps=0.0002,0.0020,0.0040,
%0.0080,0.0200,0.1800$ 
%from bottom to top.
(b)The instantaneous diffusion index $\alpha_{ins}(t)$.
The broken line indicates the critical super-diffusion line
$\alpha_{ins}(t)=\alpha_c=1.73$.
(c) The scaled MSD $\Lambda(t)=m_2(t)/t^{1.73}$
as functions of time for increasing strengths $\eps$.
The results for $\eps_b \simeq 0.10$ are shown in thick blue lines.
%The range of $\eps$ is $\eps \in [0.02,0.8]$.
%$\eps=0.17,0.18,0.19,0.20,0.21,0.22,0.23,0.25$ from below.
}
\end{center}
\end{figure}

%%%%%%%%%%%%%%%%%%%%%%%%%%%
{\it Ballistic-Diffusive transition (BDT).-}
%%%%%%%%%%%%%%%%%%%%%%%%%%%
%\red{色数$M$に就いて触れていないが：色問題 一色ではどうだったか？
%二色以上は転移があるようだが、M=3とした理由が必要ではないか？}\\
%\red{
%\blue{
Next, we study the effect of the harmonic perturbation on 
the extended states in the KHM which occurs in the case $V\ll 1$.
In such a regime the initial wave packet spreads ballistically as 
$m_2(t) \sim t^2$ without the harmonic perturbation.
In view of such dynamical behavior, we henceforth refer to 
the extended state as a ballistic state.
In the case of $M = 1$ the ballistic spreading does not change, and
we investigate the case of $M \geq 2$ 
in accordance with the case of LDT.
% as in Fig.\ref{fig:LDT}.
We take the case of $V=0.2$ 
and $M=3$ as the typical case in accordance with the case 
where LDT was observed in Fig.\ref{fig:LDT}.
%}
The time-dependence of MSD for various values of $\eps$
is displayed in Fig.\ref{fig:BDT}(a).
When $\eps$ is small, the growth of $m_2 (t)$ slightly deviates from ballistic spreading 
in the initial time regime, but it reaches to ballistic increase as $t \to \infty$. 
However as $\eps$ is taken large enough, it gradually approaches to 
normal diffusion obeying $m_2 (t) \sim t^1$
and finally it increases as $m_2 (t)\sim t$ even from the initial stage.\\ 

%%%%%%%%%%%%%%%%%%%%%%%%%%%%%%%%%%%%%%%%%%
%(a)c3n17b.pxp/c3n17/V02-A-balli-eps/sin-KHM-A-Balli-eps
%(c)c3n17b-scale.pxp/c3n17/V02-A-balli-eps/sin-KHM-A-Balli-eps
%(b)alpha-balli.pxp/alpha-Atype-KHM-BDT/sin-KHM-A-Balli-eps
%%%%%%%%%%%%%%%%%%%%%%%%%%%%%%%%%%%%%%%%%%

The above observation suggests a transition from the ballistic spreading to 
a normal diffusion. We confirm this directly by the $\alpha_{ins}(t)$ vs $t$-plot
shown in Fig.\ref{fig:BDT}(b).
%As seen in Fig.\ref{fig:BDT}(b) for $M=3$ around $\eps\simeq \eps_b (\simeq 0.10)$.
%, the presented cases are
%characterised by the superdiffusion ($1<\alpha<1$).
In all cases $\alphains(t)$ increases in the initial stage. For $\eps$
large enough the increase continues to $t\to\infty$, whereas $\alphains(t)$
decreases as $t\to\infty$ for $\eps$ small enough . There exist a
certain $\eps=\eps_b$ at which $\alphains(t)$ approaches to 
a constant value $\alpha_b\sim 1.73$, indicating the
presence of the asymptotic super-diffusion $m_2\propto t^{\alpha_c}$ 
as the critical state.
%shows a relatively stable $\alpha \simeq 1.75$
% at a certain value of $\eps_b$, and then $m_2(t) \sim t^\alpha$ is suggested 
% to transition from $\alpha \to 2$ to $\alpha \to 1$.
%As shown in Fig.\ref{fig:BDT}(c), the scaled MSD $\Lambda(\eps,t)=m_2(t)/t^{1.75}$ 
%for various $\eps$ has the trampet-shaped pattern.

\begin{figure}[htbp]
\begin{center}
\includegraphics[width=5.0cm]{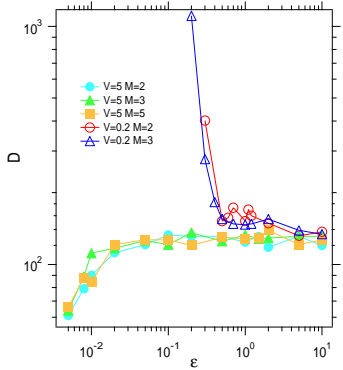}
\caption{\label{fig:Map-ABalli-Deps}(Color online)
Diffusion coefficient $D$ as a function of $\eps$ 
of  the localized  ($V=5$) and extended  ($V=0.2$) states in the  perturbed 
KHM. $\hbar=1/8$.
%Note that the both axes are in logarithmic scale.
%両軸は対数表示。
}
\end{center}
\end{figure}

Figure \ref{fig:Map-ABalli-Deps} is an estimate of the diffusion coefficient $D$ 
in the normal diffusion region ($\eps>\eps_b$). 
The time-dependence of MSD shows the normal diffusion 
for $\eps>\eps_b$  and the diffusion coefficient decreases by increasing $\eps$.
On the other hand, as seen in the previous part, for $\eps>\eps_c$
$m_2(t)$ gradually approaches to the normal diffusion for  $t \to \infty$.
Here, in the KHM the $\eps-$dependence of the diffusion coefficient in the normal diffusion region
in a wide region of $\eps$($>\max\{\eps_c, \eps_b\}$)  is observed in Fig.\ref{fig:Map-ABalli-Deps}.

%%%%%%%%%%%%%%%%%%%%%%%%%%%%%%%%%%%%%%%%%%
%Map-ABalli-Deps.pxp/Balli-Loc-Diff  hbar=1/8 theta=0
%%%%%%%%%%%%%%%%%%%%%%%%%%%%%%%%%%%%%%%%%%

%\red{以下は''Summary and Discussion''に回した方がよいかも\\
In the localized case ($V=5$), the diffusion coefficient increases with an increase of $\eps$. 
As $\eps$ is increased far beyond $\eps_c$ and the decrease saturates 
at a certain level of the order of $O (1)$. 
On the other hand, in  the extended case ($V=0.2$), the diffusion coefficient decreases monotonically 
as $\eps$ increases and for $\eps$ far beyond $\eps_b$, 
$D$ falls down to the same level as the localized cases 
for $\eps>>1$. This fact implies that for any $V$ there always exist a critical point $\eps_c$ beyond which
an irreversible diffusion switches on, whose diffusion coefficient converges to the same level.
%}

%%%%%%%%%%%%%%%%%%%%%%%%%%%
{\it Summary and discussion.-}
%%%%%%%%%%%%%%%%%%%%%%%%%%%
%\red{●$V\to 0$でも$\eps$がO(1)になれば拡散
%が起きて全く不思議でないし、必ず起こるだろう。その場合も臨海転移
%を伴うに違いない（？）。しかし$V=0.2$のBDTでも$eps$が0.1程度とい
%う小ささで転移が起こることが重要なのか？ そうも見えるが.....
%総括すれば\unl{KHMではVが小さくなるにつれ$\eps_c$が大きくなる}
%という事が言える？ ならばいまとなっては$\eps_b$をやめて$\eps_c$
%で一貫すべきだろうか？\\
%●なぜ$V$が大きいと小さい$\eps_c$で転移が起こるのか？\unl{その小ささ
%はナニできまるのか？}\\
%てえな事をこってりと議論する必要がないか？}\\
%%%%%%%%
We investigated the dynamical property 
 of the initially localized wave packet 
 in coherently perturbed KHM.
In the localized cases ($V>>1$), the LDT 
appeared with increasing the perturbation strength $\eps$. 
The critical value $M_c$ which appear the LDT is $M_c=2$.
It also describes the localized and delocalized transitions 
in the multidimensional Anderson model that corresponds to 
the 1D Anderson map and kicked rotors  with the quasi-periodic perturbation. 
If $M+1$ can be identified with the spatial dimension $d$, the 
 existence of the LDT is a qualitatively consistent result with those of 
the LDT in  $d-$dimensional Anderson model 
\cite{anderson58,abrahams79,lifshiz88,abrahams10}.
It was also shown that 
the ballistic dynamics without the perturbation
makes transition to the normal diffusion with increase in the perturbation strength.
%It is shown that a dynamical transition from ballistic to diffusive spread. 
It will be interesting to see if similar results presented in this paper 
can be obtained for the original Harper model with localized, 
critical and delocalized states as well as the KHM,

As a result, it was shown that 
whether the unperturbed state is localized or ballistic, the increase in the 
strength of the coherent harmonic perturbation causes a transition to the normal diffusion.
%Therefore, we can expect that the quasi-periodically driven system also may delocalize 
%even in the absence of coupling with its environment.
The asymptotic diffusive behavior is expected as a generic feature 
of decoherence, when noise is not introduced into a system.
%\red{この一言でことの重要さの本質を表現できないか？
The observed facts presented above implies that in order to induce quantum diffusion, which is apparently an 
irreversible quantum dynamics, we need neither spatial nor temporal stochasticity
and only spatial and temporal aperiodicity is sufficient.
%}
This work may give some hints leading to a deeper understanding of dynamical localization
and quantum diffusion in quasi-periodic systems.
%This study in the Kicked Harper model and Harper model 
It also provides insight into 
the control of localized and delocalized states by coherent perturbations in the Floquet engineering.

\vspace{1cm}
%%%%%%%%%%%%%%%%%%%%%%%%%%%%%%%%%%%%%%%%%%%%%%%
%%%%%%%%%%%%%%%%%%%%%%%%%%%%%%%%%%%%%%%%%%%%%%%
{\it Acknowledgments:}
This work is partly supported by Japanese people's tax via JPSJ KAKENHI 15H03701,
 and the authors would like to acknowledge them.
They are also very grateful to Dr. T.Tsuji and  Koike memorial
house for using the facilities during this study.
The author (H.Y.) would like to acknowledge the hospitality of the Physics Division of the 
Nippon Dental University at Niigata, where part of this work was completed.

%\appendix

\vspace{1cm}
%%%%%%%%%%%%%%%%%%%%%%%%%%%
{\it Appendix: Maryland transform.-}
%%%%%%%%%%%%%%%%%%%%%%%%%%%
We can regard the time-dependent harmonic perturbation $f_\eps(t)$ 
as the dynamical degrees of freedom. To show this
we introduce the classically canonical action-angle operators 
$(\hatJ_j=-i\hbar \frac{\pr_j}{\pr_j\phi_j}, \phi_j)$
representing the harmonic perturbation as the linear modes, and we 
 call them the color modes.
 Each quantum oscillator has the action eigenstates $|n_j>$ 
 with the action eigenvalue $J_j=n_j\hbar~(n_j:$integer) 
and the energy $n_j\hbar\omega_j$, 
where $\hatJ_j|m_j\>=m_j\hbar|m_j\>$($m_j\in {\Bbb Z}$).
Thus the system (\ref{eq:Hamiltonian}) is 
regarded as a quantum  system of $(M+1)$-degrees of freedom 
spanned by the quantum states $|n>\prod_{j=1}^M |n_j>$. 
Then  the Hamiltonian $\tilde{H}_{kick}$ that include the color modes becomes
\beq
&& \tilde{H}_{kick}(\hatp,\hatq,\{\hatJ_j\},\{\hatphi_j\}) =2\cos(\hatp/\hbar)+ \nn \\
&& 
%+ V(\hatq,\{\hatphi_i\}) \delta_t
2V\cos(2\pi Q\hatq) \left[1+ \frac{\eps}{\sqrt{M}} \sum_j^M \cos \phi_j \right]\delta_1(t)
+\sum_{j=1}^M \omega_j\hatJ_j,
\eeq
%where  $\delta_t=\sum_{m=-\infty}^{\infty}\delta(t-m)$.
where $\gamma$ and $|u\>$ are the quasi-eigenvalue and quasi-eigenstate.
Here, if the eigenstate representation of $\hatJ_j$ is used,
we can obtain the following 
$(M+1)-$dimensional tight-binding expression 
by the Maryland transform \cite{fishman82,yamada20}:
%\begin{widetext}
\beq
%\displaystyle 
\label{eq:Maryland_AM}
& & D(n,\{m_j\})u(n,\{m_j\}) +  \nn \\
& & \sum_{n',\{m_j^{'}\}}\<n,\{m_j\}|\hat{t}_{KHM}|n',\{m_j^{'}\}\>  
u(n',\{m_j^{'}\}) =0, 
\eeq
where $\{m_j\}=(m_1,....,m_M)$.
Here the diagonal term is 
\beq
D(n,\{m_j\})=
\tan \left[ \frac{2V\cos(2\pi Q n)+\hbar\sum_j^Mm_j\omega_j}{2\hbar}-\frac{\gamma}{2}
 \right], 
\label{eq:diagonal}
\eeq
and the $\hat{t}_{KHM}$ of the off-diagonal term is 
\beq
\hat{t}_{KHM}=i\frac{
e^{-i  \frac{\eps 2V}{\sqrt{M}}\cos(2\pi Q\hatq) (\sum_{j}^{M}\cos\phi_j)/\hbar}-
e^{i2\cos(\hatp/\hbar)/\hbar}
}
{e^{-i\frac{\eps 2V}{\sqrt{M}}\cos(2\pi Q\hatq) (\sum_{j}^{M}\cos\phi_j) /\hbar}+
e^{i2\cos(\hatp/\hbar)/\hbar}
}.
\eeq
%\blue{
The effect of $V$ in the on-site-potential $D(n,\{m_j\})$ is saturated for $|V/\hbar|>>\pi$,
since the aperiodicity is expressed through the tangent function.
The off-diagonal component controlling inter-site-hopping is governed 
by a single parameter $\eps V/\hbar$.
This is consistent with the fact that when $|V|$ is sufficiently large,
 it is the combined parameter $\eps V$ (more precisely $\eps V/\hbar$) that 
 governs the feature of the transition.
%}

%\red
%{○池田\\
%「on-site-potentialのaperiodicityはtan関数を通して表現されているので$|V/\hbar|$
%が$\pi$を越すほど$|V|$が大きければ、その効果は飽和してしまう。そしてinter-site
%hoppingをコントロールするoff-diagonal成分は$\eps V/\hbar$というsingle parameter
%で支配されている。このことは$|V|$が十分に大きい時、転移のfeatureを支配するのは
%は combined parameter $\eps V$ (正確には$\eps V/\hbar$) であることとconsistent。」と言いたい？
%}\\

%%%%%%%%%%%%%%%%%%%%%%%%%
%%%%%%%%%%%%%%%%%%%%%%%%%

%%%%%%%%%%%%%%%%%
%%%%%%%%%%%%%%%%%
\end{document}